\documentclass[conference,10pt,letter]{IEEEtran}                         
\pdfoutput=1
\IEEEoverridecommandlockouts                              

\pdfpageattr{/Group << /S /Transparency /I true /CS /DeviceRGB >>}

\usepackage[utf8]{inputenc}
\usepackage[T1]{fontenc}
\usepackage{textcomp}
\usepackage{fixltx2e}
\usepackage{microtype}
\usepackage{calc}

\usepackage[range-phrase=--,per-mode=symbol-or-fraction,binary-units=true,range-units=single,list-units=single,detect-all]{siunitx}
\usepackage{silence}
\DeclareSIUnit\bit{bit}
\DeclareSIUnit\byte{Byte}
\DeclareSIUnit\decibeli{dBi}
\DeclareSIUnit\decibelm{dBm}
\DeclareSIUnit\mph{mph}
\DeclareSIUnit\resourceblock{RB}
\DeclareSIUnit\vehicle{veh}
\DeclareSIUnit\watthour{Wh}

\usepackage{csquotes}
\usepackage[backend=bibtex,style=ieee,doi=false,isbn=false,mincitenames=1,maxcitenames=2]{biblatex}
\DeclareFieldFormat{sentencecase}{#1} 
\DeclareFieldFormat{titlecase}{#1} 
\usepackage{xpatch}
\xpatchbibmacro{textcite}{\addspace}{\addnbspace}{}{}
\xpatchbibmacro{Textcite}{\addspace}{\addnbspace}{}{}

\DefineBibliographyStrings{english}{
	andothers = et~al\adddot\addspace
}

\usepackage{amsmath}

\usepackage{amssymb}
\usepackage{amsfonts}

\DeclareMathOperator*{\argmin}{argmin}

\usepackage{booktabs}
\usepackage{url}

\usepackage[caption=false,font=footnotesize]{subfig}
\usepackage[pdftex]{graphicx}
\DeclareGraphicsExtensions{.pdf,.png,.jpg}
\usepackage{tikz}
\usetikzlibrary{calc}
\usetikzlibrary{shapes.geometric}
\usetikzlibrary{spy}

\usepackage[american]{babel}
\usepackage{hyphenat}

\usepackage{algorithmic}
\usepackage{algorithm}

\usepackage[capitalize,noabbrev]{cleveref}

\crefname{algocf}{Algorithm}{Algorithms}
\Crefname{algocf}{Algorithm}{Algorithms}

\usepackage{pgfplots}
\usepackage{pgfplotstable}
\pgfplotsset{compat=1.15}

\usepackage{multirow}
\usepackage{cancel}

\usepackage{xstring}
\usepackage{xparse}
\usepackage[nolist]{acronym}

\acrodef{CSP}{constrained shortest path}
\acrodef{EVCSP}{electric vehicle constrained shortest path}
\acrodef{EVCAS}{electric vehicle continuous adaptive speeds shortest path}
\acrodef{SOC}{state of charge}
\acrodef{ITS}{intelligent transportation systems}
\acrodef{CSDB}{charging station database}

\usepackage[color]{changebar}
\def\todoCtd#1{%
	TODO: #1%
	\ifx&#1&...\fi%
	\endgroup
	\cbend
	\relax
}

\NewDocumentCommand\IEEE{ s m d[] }{%
	\IfBooleanTF{#1}{}{IEEE\,}
	\nolinebreak[2]
	#2%
	\IfNoValueTF{#3}{%
	}{%
		\StrGobbleLeft{#3}{1}[\sommerIEEEFirstLetter]%
		\IfEq{\sommerIEEEFirstLetter}{}{%
			#3
		}{%
			\nolinebreak[3]
			\StrLeft{#3}{1}%
			\sommerIEEELettersSlashed{\sommerIEEEFirstLetter}%
		}%
	}%
}
\newcommand{\sommerIEEELettersSlashed}[1]{%
	/
	\StrLeft{#1}{1}%
	\StrGobbleLeft{#1}{1}[\sommerIEEESubsequentLetter]%
	\IfEq{\sommerIEEESubsequentLetter}{}{%
	}{%
		\sommerIEEELettersSlashed{\sommerIEEESubsequentLetter}
	}%
}

\begin{document}
\title{Reducing Waiting Times at Charging Stations with Adaptive Electric Vehicle Route Planning}
\author{Sven Schoenberg and Falko Dressler
\thanks{Research reported in this paper was conducted in part in the context of the Hy-Nets4all project, supported by the European Regional Development Fund (EFRE-0801678).}
\thanks{Sven Schoenberg is with the Software Innovation Lab and the Department of Computer Science, Paderborn University, Germany. E-Mail: sven.schoenberg@c-lab.de}
\thanks{Falko Dressler is with the School of Electrical Engineering and Computer Science, TU Berlin, Germany. E-Mail: dressler@ccs-labs.org}
}

\maketitle
\thispagestyle{empty}
\pagestyle{empty}

\begin{abstract}
Electric vehicles are becoming more popular all over the world.
With increasing battery capacities and a growing fast-charging infrastructure, they are becoming suitable for long distance travel.
However, queues at charging stations could lead to long waiting times, making efficient route planning even more important.
In general, optimal multi-objective route planning is extremely computationally expensive.
We propose an adaptive charging and routing strategy, which considers driving, waiting, and charging time. 
For this, we developed a multi-criterion shortest-path search algorithm using contraction hierarchies.
To further reduce the computational effort, we precompute shortest-path trees between the known locations of the charging stations.
We propose a central \ac{CSDB} that helps estimating waiting times at charging stations ahead of time.
This enables our adaptive charging and routing strategy to reduce these waiting times.
In an extensive set of simulation experiments, we demonstrate the advantages of our concept, which reduces average waiting times at charging stations by up to \SI{97}{\percent}.
Even if only a subset of the cars uses the \ac{CSDB} approach, we can substantially reduce waiting times and thereby the total travel time of electric vehicles.
\end{abstract}

\acresetall
\IEEEpeerreviewmaketitle

%

\section{Introduction}
\IEEEPARstart{T}{he}
capacity of batteries used by electric vehicles is continuously increasing and the trend is projected to continue for the coming years~\cite{iea2020global}.
Together with a growing fast-charging infrastructure, this makes electric vehicles more suitable for long distance travel.
Nevertheless, if recharging on a trip is necessary, the charging times can still be quite long and as more electric vehicles utilize the charging infrastructure, long waiting times are becoming a main challenge~\cite{berman2019rapid}.
There is a substantial heterogeneity in the charging infrastructure, for example with respect to available charging power, location, and potential waiting time.
Thus, the total travel time of the trip can be significantly impacted by the selection of charging stations along the trip.
We therefore need solutions helping to plan long distance trips and coordinating the use of charging stations between vehicles.


In general, it has been discovered that, due to the need to take recharging into account, route planning for electric vehicles is more challenging compared to route planning for conventional vehicles~\cite{alizadeh2014optimized, liu2019electric, baum2017consumption}.
The recharging time of a vehicle depends on multiple factors.
Most importantly, these are the amount to recharge, which depends on the energy consumed while driving, and the power of the charging station.
The driver has the option to take faster routes or more economic routes, i.e. routes that require less energy.
In order to minimize the total travel time, the driver could take a more economic route to save charging time -- at the cost of a longer driving time.
The total travel time is also influenced by the concrete amount of energy to be charged at the charging station.
This is particularly complicating the optimization problem as the charging process is not linear: 
Lithium-ion batteries, as used in most electric vehicles, are charged with a charging protocol that decreases the charging speed considerably after the battery capacity reaches about \SI{80}{\percent}~\cite{ng2009enhanced}.

Waiting times due to fully occupied charging stations can be a substantial part of the total travel time as well~\cite{deweerdt2016intentionaware}.
If the potential waiting time at each charging station is known ahead of time, it can be taken into account when planning the route.
It might take less time to take a detour to a different charging station, if the waiting time there is substantially smaller.
We also think it would make sense to take a slower, more energy efficient route, if it is known that the car has to wait at the charging station anyhow.

To solve the problem of finding a route that minimizes the total travel time for an electric vehicle, we have to select the charging stations, the amount of energy to charge at each charging station, and a route to get to and from the charging stations.
The charging station selection must take into account the driving time, the potential waiting time and the charging time.
The driving time, as well as the charging time depends on the selected route to the charging station.
These aspects also mutually influence each other.

Our solution is to use a multi-criterion shortest-path search for the criteria energy and time, revealing all Pareto-optimal routes from the most economic to the fastest route.
In a second step, the alternatives are compared to identify the best candidate.
The key challenge is that such a multi-criterion shortest-path search is rather expensive with respect to required computational resources~\cite{goodrich2014twophase}.
In our previous work~\cite{schoenberg2019planning}, we presented an approach to accelerate the search by focusing on the most popular queries, which are between the (known) locations of the charging stations.
We selected the charge amount and the route between charging stations with an adaptive charging and routing strategy.

In this paper, we go one step further and study a way for vehicles to coordinate their charging station visits to reduce waiting times.
We propose a central \ac{CSDB}, to which the vehicles announce their planned charge stops and which, in return, can estimate waiting times at charging stations.
The estimated waiting times can easily be included in our routing algorithm with the adaptive charging and routing strategy to minimize the total travel time.
Using the database is not mandatory, but it provides significant benefits to its users even if only a small subset of all vehicles uses it. 

Our main contributions can be summarized as follows:
\begin{itemize}
\item we present an improved approach (extending our prior work in~\cite{schoenberg2019planning}) for electric vehicles to coordinate their (planned) charging station visits to reduce the waiting time (\cref{section:csdb}); and
\item we perform a simulation study to analyze the effect of different penetration rates of the \ac{CSDB} and of using historical data on the travel times to study our adaptive charging and routing strategy compared to related strategies.
We show that by using the \ac{CSDB} we can reduce the average waiting time up to \SI{97}{\percent} (\cref{section:performanceeval}).
\end{itemize}

%

\section{Related Work}

The classic route planning in a road traffic network is a shortest path problem, where the best route to be found from A to B is based on some criterion.
Typically, the criterion to optimize for is either driving distance (shortest route), travel time (fastest route), energy consumption (economic route), or a combination thereof.
The best known solution is Dijkstra's algorithm~\cite{dijkstra1959note}.
%
Depending on the size of the graph, Dijkstra's algorithm can be too slow in practice, but several techniques have been proposed to speed things up.
For example, the A* algorithm~\cite{hart1968formal} uses a heuristic for a directed search.
If the heuristic is guaranteed not to overestimate the cost, A* will find the optimal path.

Another technique is the use of contraction hierarchies introduced by \textcite{geisberger2008contraction}.
In a preprocessing step, shortcuts are added to the graph that can later speed-up the path finding query significantly.
This is done by contracting the nodes of the graph one by one.
Each node that is contracted, is effectively removed from the graph.
If the node was part of the shortest path between two of its neighbors, a direct edge between these neighbors (shortcut) is added to ensure that the shortest path is maintained.
Whether this is the case can be determined by doing a shortest path search with Dijkstra's algorithm from each neighbor to all other neighbors.
Each node is assigned a level based on the order of contraction.
A higher level indicates that the node was contracted later and its shortcuts might have replaced shortcuts of lower level nodes.
To query the shortest path, a bidirectional search with Dijkstra's algorithm is done with both sides only traversing to nodes that have a higher level until they meet.
This way, the number of nodes that need to be visited to find the shortest path is reduced significantly.
%
To further speed-up the query, A* can be used instead of Dijkstra's algorithm for the bidirectional search~\cite{baum2020modeling}.

Route planning for electric cars presents additional challenges.
The constraints of the battery, especially the limited range, have to be accounted for.
This can further include recuperation, also called regenerative braking, i.e., charging the battery when slowing down or driving downhill.
Finding the shortest path that also considers such battery constraints is a \ac{CSP} problem~\cite{artmeier2010shortest}.
To find the fastest route that is reachable with a limited range, a multi-criterion shortest path search can be performed using the criteria travel time and energy consumption.
This results in all Pareto optimal paths for these criteria and we can, for example, choose the one with the best travel time that still fulfills the energy constraints.
To calculate all Pareto optimal paths, a modified version of either Dijkstra's algorithm or the A* algorithm can be used; however, due to the increased complexity, this is even less practical than for a single criterion.
Fortunately, contractions hierarchies can also be used to speed-up multi-criteria path finding and to solve the \ac{CSP} in acceptable time~\cite{storandt2012quick}.

The preprocessing step for contractions hierarchies for multi-criteria path finding requires significant computational effort for large (e.g., country-sized) maps.
As more and more nodes are contracted, the remaining uncontracted nodes get more and more neighbors, which makes contracting the last few nodes very expensive.
It is possible to restrict the preprocessing to only contract a subset of all nodes.
For example, \textcite{storandt2012route} contracted only \SI{99.5}{\percent} of the nodes to achieve reasonable preprocessing times.
The remaining uncontracted graph is called a core graph~\cite{dibbelt2012userconstrained}.
This can substantially save preprocessing time, but also causes higher query times~\cite{schoenberg2019planning}.

Another approach by \textcite{baum2020modeling} introduces a way to solve the \ac{EVCAS} problem, which they define as finding a feasible path that respects battery constraints, considers variable speed and minimizes driving time.
To speed-up queries, they use contraction hierarchies together with A* and can compute optimal solutions in less than a second even for large battery sizes and country sized maps.
For interactive applications, very low query times are more important than getting optimal solutions.
By using a heuristic, query times can be improved significantly at the cost of some inaccuracy~\cite{baum2020modeling, hartmann2014energyefficient}.

The problem becomes more complicated when recharging on the way is also considered, to enable long distance trips.
One solution is to limit the number of recharging events and choosing the most economic route that is not more than \SI{10}{\percent} longer than the shortest route~\cite{storandt2012quick}.
\Textcite{morlock2019time} solve the problem by first querying a number of potential routes ranging from shortest to fastest from a conventional routing service and treating the routes as a reduced graph.
They then integrated the charging stations into the graph and performed a multi-objective shortest path search to find the fastest route from start to destination.
Most publications assume a full recharge at each charging station~\cite{storandt2012quick, goodrich2014twophase, sun2016save, deweerdt2016intentionaware, morlock2019time}, only few also consider partial charging~\cite{baum2017consumption, alizadeh2014optimized}.

One assumption in these works is that they consider only one vehicle and assume it can immediately start charging when it arrives at a charging station.
In reality, there will often be multiple vehicles that want to charge at the same time, but the number of vehicles that can charge concurrently is limited by the number of charge points of the charging station.
If there is no coordination between the vehicles, this will lead to queues and long waiting times.
A possible solution to this problem is a reservation system, where the vehicles can reserve a time slot at a charging station in the future and can therefore plan their trip accordingly to avoid waiting times.
Many publications make use of such a reservation system~\cite{kim2010efficient, qin2011charging, bedogni2016route, cao2019electric, hou2020bidding, lee2020deep}, sometimes with the possibility to update the reservation, if needed~\cite{qin2011charging}.
Most of these systems have a first-come-first-serve policy, but some can also prioritize reservations, leading to cases where high priority vehicles can charge before others even if they arrive later at the charging station~\cite{cao2019electric}.
\Textcite{hou2020bidding} use a scheduler that allocates reservations based on user given information about their time preferences.
Due to the assumption that users are selfish and do not want to reveal their true time preferences to avoid unfavorable time slots, they propose an iterative auction which, by progressively eliciting the users' preferences as necessary, preserves their privacy.

Recently, some approaches for charging station and route selection have been using deep reinforcement learning.
It enables them to make complex decisions in a stochastic environment with changing conditions like traffic, weather, dynamic charge prices etc. by learning an optimal policy.
\textcite{qian2020deep} present a charging navigation solution which aims to minimize the total travel time and charging cost.
It can take into account waiting times at charging stations, traffic conditions and charge prices, thereby coordinating smart grid and intelligent transportation systems.
However, they do not consider direct coordination between vehicles, but simply assume that charging stations know how long the waiting times will be.
\textcite{lee2020deep} propose a similar system where there is coordination between vehicles with a reservation system and charging decisions are made by a central service.
However, both solutions suffer from poor scalability.
They evaluated very small instances with graphs of only 39 nodes and three charging stations.
\textcite{zhang2021effective} use deep reinforcement learning for planning charging scheduling at a larger scale.
They evaluated instances of a big city with more than 1000 charging stations.
However, they use Dijkstras's algorithm for route selection and a simple energy consumption model, which only depends on the driven distance.
While this may be sufficient for inner city navigation, more sophisicated models are needed for long distance navigation.

A different approach is a centralized service that knows about the current charging station utilization and can give vehicles advice on where to charge~\cite{yang2013charge}.
The vehicles could also announce their charging intentions to this service, so that it can predict the waiting time in the future.
\Textcite{deweerdt2016intentionaware} call this intention-aware routing.
They combined the information about charging intentions with historical data and were able to reduce waiting times in some cases by about \SI{80}{\percent}.
However, they do not consider long trips with multiple charge stops or partial charging and assume charge stops always take a fixed time.

We go one step further and combine adaptive charging and routing strategies for long trips with a centralized service that can estimate waiting times at charging stations with the current utilization, planned charge stops of other vehicles, and historical data.
In contrast to a reservation system, it does not require cooperation from charging station providers and is not mandatory for the vehicles to use.
In addition, we also make use of a realistic nonlinear charging model to consider partial charging of the batteries.

%

\section{Charging Station Routing}
\label{section:chargingstationrouting}
This section describes our initial approach~\cite{schoenberg2019planning} for the ease of the reader to understand our novel concept presented in~\cref{section:csdb}.

The goal of our charging station routing approach is to minimize the total travel time of an electric vehicle on a long distance trip.
This includes finding a route from the origin to the destination and selecting charging stations along the way to recharge the vehicle if necessary.
The total travel time consists of the driving time as well as the waiting and charging time at the charging stations.
The selection of the charging stations has a big influence on the route, because the consumed energy while driving has to be recharged.
Depending on the charge power of the charging station this can lead to significant charging time.
Choosing a slower but more economical route might be faster overall if it reduces the charging time at the charging station enough.

Taking all this into account, our approach consists of two parts.
The first part is an approach for multi-criterion shortest path finding that accelerates the search between the known locations of the charging stations.
It enables us to query all Pareto-optimal routes from the most economic route to the fastest route. 

The second part is the selection of charging stations and routes that minimize the total travel time.
We compare the route alternatives, returned by the multi-criterion shortest path finding, between charging stations and the origin and destination.
Our adaptive charging and routing strategy then selects the optimal combination of charging stations and routes. 

\subsection{Multi-Criterion Shortest Path Finding}
\label{subsection:multicriterionshortestpathfinding}

Dijkstra's algorithm can be modified to be used for multi-criterion shortest path finding.
Instead of setting a single label per node to denote the predecessor node and the minimum cost to this node, we maintain a Pareto set of labels at each node.
A label contains the costs for all criteria and the predecessor node.
The path finding query is a lot more computationally expensive than for a single-criterion shortest path search.
The nodes have to be visited multiple times and new labels have to be checked with existing labels for dominance to maintain the Pareto set.

We use contraction hierarchies to accelerate the path finding.
Even though this improves the query times significantly, they might still be in the order of seconds or even minutes, especially for long distances of \SI{>200}{\kilo\meter} on a dense street network.
Using only this, the total computation time of our algorithm would be unacceptable, because it makes many queries between the origin, destination and the charging stations.

\subsection{Shortest-Path Tree Precomputing}
\label{subsection:shortestpathtreeprecomputing}

Most of the query time is used to explore the graph and maintain the Pareto sets of labels at the nodes.
Our algorithm queries only the routes between the charging stations and the origin and destination.
Because the charging station locations are always the same, we would explore the graph from the same nodes over and over again.
We can avoid this, by exploring the graph from all locations of the charging stations once in a preprocessing step.
The result of an exploration is a shortest-path tree, which contains the sets of labels of all explored nodes.
The exploration can be limited to an energy consumption equal to the maximum battery capacity of the vehicles.
Because the query of contraction hierarchies is bidirectional, we have to create the shortest-path trees twice for each charging station: Once exploring the graph forwards, and once exploring backwards.
Precomputing the shortet-path trees for all charging stations is only feasible because the contraction hierarchies significantly reduce the number of nodes that have to be explored.

To perform a query with the shortest-path trees, we use the forward explored tree of the origin node and backward explored tree of the destination node.
We then identify the nodes that are covered by both trees, i.e., nodes with Pareto sets of labels in both trees.
We create the sumset of both Pareto sets for each of these nodes and remove all dominated elements.
Each set contains the costs of all shortest-paths from origin to destination via that node.
To get the costs of all Pareto optimal shortest-paths from origin to destination we combine the sets of all nodes and again remove all dominated elements.

\tikzset{font=\footnotesize}
\begin{figure}
	\centering
	\subfloat[Shortest-path tree with Pareto sets of labels]{
		\begin{tikzpicture}[every path/.style={>=latex}]
		
			\node (A) at (0,  0) [circle, fill=black, text width=4pt, inner sep=0pt, label=left:$A$] {};
			\node (B) at (2,  1) [circle, fill=black, text width=4pt, inner sep=0pt, label=above:$B$] {};
			\node (C) at (4,  1) [circle, fill=black, text width=4pt, inner sep=0pt, label=below:$C$] {};
			\node (D) at (6,  1) [circle, fill=black, text width=4pt, inner sep=0pt, label=right:$D$] {};
			\node (E) at (3, -1) [circle, fill=black, text width=4pt, inner sep=0pt, label=above:$E$] {};
			
			\draw[->] (A) to[bend left=20]  node[above] {$(1, 2)$} (B);
			\draw[->] (A) to[bend right=20] node[above] {$(4, 1)$} (C);
			\draw[->] (A) to                node[below] {$(5, 2)$} (E);
			\draw[->] (B) to[bend left=45]  node[above] {$(3, 4)$} (D);
			\draw[->] (B) to                node[above] {$(1, 3)$} (C);
			\draw[->] (C) to                node[above] {$(1, 2)$} (D);
			
			\node at (A) [below, outer sep=4pt] {\includegraphics{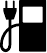}};
			
			\node at (B) [draw, above, outer sep=16pt] {(1,2,A)};
			\node at (C) [draw, below, outer sep=16pt, align=left] {(2,5,B)\\(4,1,A)};
			\node at (D) [draw, below, outer sep=6pt, align=left] {(3,7,C)\\(4,6,B)\\(5,3,C)};
			\node at (E) [draw, below, outer sep=6pt] {(5,2,A)};
		
		\end{tikzpicture}
	}
\vspace{0.01em}
	\subfloat[Forward and backward shortest-path trees with labels of common nodes. Solid: Forward tree from node A. Dashed: Backward tree from node G] {
		\begin{tikzpicture}[every path/.style={>=latex}]
		
			\node (A) at (0, 0) [circle, fill=black, text width=4pt, inner sep=0pt, label=left:$A$] {};
			\node (B) at (1, 1) [circle, fill=black, text width=4pt, inner sep=0pt, label=above:$B$] {};
			\node (C) at (2, 1) [circle, fill=black, text width=4pt, inner sep=0pt, label=above:$C$] {};
			\node (D) at (3.5, 1) [circle, fill=black, text width=4pt, inner sep=0pt, label=above:$D$] {};
			\node (E) at (1.5, -1) [circle, fill=black, text width=4pt, inner sep=0pt, label=above:$E$] {};
			
			\draw[->] (A) to[bend left=20]  (B);
			\draw[->] (A) to[bend right=20] (C);
			\draw[->] (A) to                (E);
			\draw[->] (B) to[bend left=45]  (D);
			\draw[->] (B) to                (C);
			\draw[->] (C) to                (D);
			
			\node at (A) [below, outer sep=4pt] {\includegraphics{figures/chargingstation}};
			\node at (D) [draw, below left, outer sep=3pt, align=left] {(3,7,C)\\(4,6,B)\\(5,3,C)};
			\node at (E) [draw, below left, outer sep=3pt] {(5,2,A)};
		
			\node (F) at (6,  0) [circle, fill=black, text width=4pt, inner sep=0pt, label=left:$F$] {};
			\node (G) at (7,  0) [circle, fill=black, text width=4pt, inner sep=0pt, label=right:$G$] {};
		
			\draw[<-, densely dashed] (G) to[bend right=45] (D);
			\draw[<-, densely dashed] (G) to                (F);
			\draw[<-, densely dashed] (G) to                (F);
	 		\draw[<-, densely dashed] (F) to[bend right=35] (D);
			\draw[<-, densely dashed] (F) to[bend left=15]  (E);
	
			\node at (G) [below, outer sep=4pt] {\includegraphics{figures/chargingstation}};
			\node at (D) [draw, below right, outer sep=3pt, align=left, densely dashed] {(1,6,G)\\(3,5,F)};
			\node at (E) [draw, below right, outer sep=3pt, densely dashed] {(3,4,F)};
		\end{tikzpicture}
	}
\vspace{0.01em}
	\subfloat[Create sumsets from node labels and combining them to get resulting Pareto set] {
		\begin{tikzpicture}[every path/.style={>=latex}]
			\node (DA) at (0, 1) [draw, above, outer sep=2pt, align=left] {(3,7,C)\\(4,6,B)\\(5,3,C)};
			\node (DB) at (0, 1) [draw, below, outer sep=2pt, align=left, densely dashed] {(1,6,G)\\(3,5,F)};
			
			\node (EA) at (0, -1) [draw, above, outer sep=2pt] {(5,2,A)};
			\node (EB) at (0, -1) [draw, below, outer sep=2pt, densely dashed] {(3,4,F)};
			
			\node (DC) at (3, 1) [draw, align=left] {
				(4,13,C,G)\\
				\cancel{(6,12,C,F)}\\
				(5,12,B,G)\\
				\cancel{(7,11,B,F)}\\
				(6,9,C,G)\\
				(8,8,C,F)
			};
			
			\node (EC) at (3, -1) [draw] {(8,6,A,F)};
			
			\node (RES) at (6, 0) [draw, align=left] {
				(4,13,C,G,D)\\
				(5,12,B,G,D)\\
				(6,9,C,G,D)\\
				\cancel{(8,8,C,F,D)}\\
				(8,6,A,F,E)};
			
			\draw[->] (DA) to (DC);
			\draw[->] (DB) to (DC);
			\draw[->] (EA) to (EC);
			\draw[->] (EB) to (EC);
			\draw[->] (DC) to (RES);
			\draw[->] (EC) to (RES);
			
		\end{tikzpicture}
	}
	\caption{Example query with precomputed shortest-path trees}
	\label{fig:shortestpathtree}
\end{figure}

An example of such a query with shortest-path trees for two criteria is depicted in \cref{fig:shortestpathtree}.
Each label contains the costs for both criteria and the predecessor node, which is needed to reconstruct the path later.
The common nodes, which are covered by both trees, are identified and the sumsets for these nodes are created.
The elements of the sumset contain the sum of the costs as well as the predecessor nodes of both labels.
Then, the elements of all sumsets are combined into one Pareto set, with each element also storing the node of the sumset.  
This set contains the costs of all Pareto optimal paths and the information necessary to reconstruct them.

Our charging station routing algorithm makes queries between the charging stations and the origin and destination.
With the shortest-path trees precomputed for all charging stations, we just need to create the shortest-path trees for the origin and destination and can then answer all queries without exploring the graph again.
We do not have to reconstruct all paths; in order to save time, we can select a path based on the costs and only reconstruct the selected one.

\subsection{Adaptive Charging and Routing Strategy}
\label{subsection:chargingandroutingstrategy}

Our adaptive charging and routing strategy tries to find a route from the origin to the destination that minimizes the total travel time.
If necessary, it selects charging stations along the way and determines the amount of energy to recharge.
The selection of the charging stations is done with a shortest-path search on a dynamically generated graph that connects origin, destination and charging station nodes that are reachable within the vehicles range.
We use an A* search to perform the shortest-path search on the graph.
The edge weights are generated by querying all Pareto optimal paths between the two nodes and selecting the one with the lowest sum of charging time, driving time and waiting time.
The heuristic is not the linear distance to the destination, but determined with a single criterion shortest-path search.

The charging time is calculated by our charging model (cf.\ \cref{section:chargingmodel}) and depends on the power of the charging station as well as the amount of energy to recharge.
Many publications assume a full charge at every charge stop.
Our adaptive charging strategy selects the amount to recharge based our charging model and the maximum charge power of the next charging station.
We continue to charge as long as the maximum charge power of the next charging station is below the current charge power.
We, of course, always charge at least the energy required to be able to reach the next charging station.

In our dynamically generated graph, each edge weigth contains the charging time at the current charging station, the driving time to the next charging station, and also the expected waiting time and charging time at the next charging station.
We include the charging time at the next charging station, so that the route selection is influenced by how much time it takes to recharge the spent energy. 
Because we do not know at this point how much we will charge at the next charging station, we temporarily assume to fully recharge.
The time for the full recharge will be replaced by the actual charging time that is determined when calculating the next edge weight.
While exploring the graph, we propagate the batteries \ac{SOC}.

Shortest-path tree precomputing reduces the query time significantly, but it is still orders of magnitudes slower than a single-criterion shortest-path search.
To keep the number of multi-criterion queries as low as possible, when we explore a node and have to calculate the edge weights to all neighbors, we set it to a temporary heuristic value based on two single-criterion shortest-path searches for the fastest and most energy efficient routes.
Only when the edge is about to be travelled as part of the A* graph exploration, because the corresponding node is at the top of the open list, do we replace the value with the accurate value from a multi-criterion shortest-path search.
It might happen that the node is then no longer at the top of the open list, in which case the process has to be repeated.

%

\section{Charging Station Database (CSDB)}
\label{section:csdb}

\begin{figure}
	\centering
	\begin{tikzpicture}[every path/.style={>=latex}]
		\node (A) at (0,  0) [cylinder, draw, shape border rotate=90, inner sep=7pt, outer sep=5pt, text width=14pt, aspect=0.3] {CS\\DB};
		\node (B) at (-3,  -3) [inner sep=0pt, outer sep=5pt, label=below:Electric vehicle] {\includegraphics{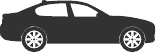}};
		\node (C) at (3,  -3) [inner sep=0pt, outer sep=5pt, label=below:Charging station] {\includegraphics{figures/chargingstation}};
	
	 	\draw[->] (A) to[bend right=20] node[left, text width=50pt, align=center] {waiting times} (B);
	 	\draw[->] (B) to[bend right=20] node[right, text width=60pt, align=center] {planned\\charge stops} (A);
	 	\draw[->] (C) to                node[right, outer sep=5pt, text width=60pt, align=center] {current utilization} (A);
	 	\draw[-]  (B) to                node[below] {charging} (C);
	\end{tikzpicture}
	\caption{Charging station database concept}
	\label{fig:csdboverview}
\end{figure}

To coordinate charging station visits between the electric vehicles, we propose the use of a centralized \ac{CSDB} (please note that our algorithm can also be executed using mobile edge computing~\cite{dressler2019virtual}).
It can estimate waiting times at charging stations in the future, which can be used by the vehicles when they calculate their route.
To be able to do this, the vehicles have to announce their planned charge stops to the database.
It also gets information about the current utilization of the charging stations and stores statistical data about past utilization.
The principle of our \ac{CSDB} approach is depicted in \cref{fig:csdboverview}.

With the adaptive charging and routing strategy, the charging station routing algorithm can easily take advantage of the waiting time information as an additional time cost at each charge stop.
For instance, if a vehicle will have to wait at a charging station anyway, it might as well drive a slower but more energy efficient route.
Even though it arrives later, it will start charging at the same time and the reduced energy consumption saves charging time, improving the overall total travel time.
Another example might be to make a detour to a different charging station if the saved waiting time is greater than the additional driving time.

Compared to a reservation system, we believe it is far more practical, because there cannot be a situation where reserved but empty charge points lead to reduced average utilization and bad experience from drivers not using the system.
Another advantage is, that it does not require cooperation with the charging station providers.
Thus, only information about the current utilization of the charging stations is needed, which many providers already provide as a service to potential customers.
The system also does not require every vehicle to take part in it to be useful.


\begin{table}
	\setlength{\extrarowheight}{.2em}
	\caption{Description of symbols}
	\centering
	\begin{tabular}{ll}
	\toprule
 	Symbol & Description\\
 	\midrule
	$S$				& Set of charging stations\\
	$C_s$			& Set of charge points of charging station $s$\\
	$t_0$			& Query time\\
	$t_{arr}^q$		& Queried arrival time\\
	$t_{wait}^q$	& Resulting waiting time of query\\
	$t_{start}^q$	& Resulting charge start time of query\\
	$P_s$			& Set of planned charge stops of charging station $s$\\
	$t_{arr}^p$		& Arrival time of planned charge stop $p$\\
	$t_{start}^p$	& Charge start time of planned charge stop $p$\\
	$t_{chr}^p$		& Charging time of planned charge stop $p$\\
	$t_{dep}^p$		& Departure time of planned charge stop $p$\\
	$c_p$			& Charge point assigned to planned charge stop $p$\\
	$T$				& Period of charge stops for statistical utilization\\
	$t_{chr}$		& Charging time of charge stops for statistical utilization\\
	$u$				& Statistical utilization of charging station\\
	$n$				& Number of charge points of charging station\\
	\bottomrule
	\end{tabular}
	\label{tab:symboldescription}
\end{table}

The main service that the \ac{CSDB} provides is estimating waiting times at charging stations for an arrival time in the future.
The estimate is created by combining data about the current utilization of the charging station, planned charge stops of other vehicles, and the statistical utilization of the charging station.
Vehicles can query estimates for any charging station and any point of time in the future.
To ease the reading of the following part, we provide a list of the used symbols in \cref{tab:symboldescription}.

The current utilization of a charging station is known to the database in the form of occupied charge points and the time when the vehicles occupying the charge points will depart.
For each charge point $c \in C_s$ of a charging station $s \in S$, we denote the departure time of the occupying vehicle as 
$t_{dep}^{c}$. 
In case the charge point is not occupied by a vehicle, we define $t_{dep}^{c} = t_0$, with $t_0$ being the time the query is made.

The database also contains planned charge stops $P$ for all charging stations.
Each of the planned charge stops $p \in P$ consists of an announced arrival time $t_{arr}^p$ and charging time $t_{chr}^p$.
The charging start time $t_{start}^p$ depends on the arrival time at the charging station (cf.\ \cref{eq:tstart}).
If we add the charging time, we get the departure time $t_{dep}^p$:
\begin{equation}
	t_{start}^p = t_{start}(t_{arr}^p) \text{ ,}
\end{equation}
\begin{equation}
	t_{dep}^p = t_{start}^p + t_{chr}^p \text{ .}
\end{equation}
The planned charge stop is assigned to the charge point $c \in C_s$, which, regarding the arrival time, would be free soonest:
\begin{equation}
	c_p = \argmin_{c \in C_s}(t_{free}^c(t_{arr}^p)) \text{ .}
\end{equation}
We denote planned charge stops assigned to the charge point $c$ with an arrival time earlier than the arrival time $t_{arr}$ as:
\begin{equation}
	P_c(t_{arr}) = p \in P_s, c_p = c, t_0 < t_{arr}^p < t_{arr} \text{ .}
\end{equation}
We can now define the time a charge point becomes free for an arrival time $t_{arr}$ as the last departure time of these planned charge stops or, in case there are none, the departure time of the vehicle currently occupying it as
\begin{equation}
		t_{free}^c(t_{arr}) = \begin{cases} 
			\max_{p \in P_c(t_{arr})} t_{dep}^p &\text{ if} P_c(t_{arr}) \neq \varnothing \\
			t_{dep}^c &\text{ else}
		\end{cases} \text{ .}
	\label{eq:tfree}
\end{equation}
The charging start time is the soonest time a charge point becomes free, but cannot be before the arrival time:
\begin{equation}
	t_{start}(t_{arr}) = \max(t_{arr}, \min_{c \in C_s}(t_{free}^c(t_{arr}))) \text{ .}
	\label{eq:tstart}
\end{equation}
For a queried arrival time $t_{arr}^q$, we can then simply calculate the charging start time $t_{start}^q$ and thereby the  waiting time $t_{wait}^q$:
\begin{equation}
	t_{start}^q = t_{start}(t_{arr}^q) \text{ ,}
\end{equation}
\begin{equation}
	t_{wait}^q = t_{start}^q - t_{arr}^q \text{ .}
\end{equation}

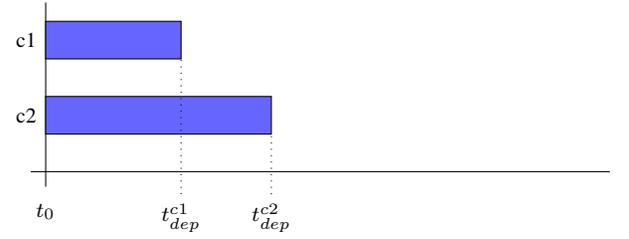
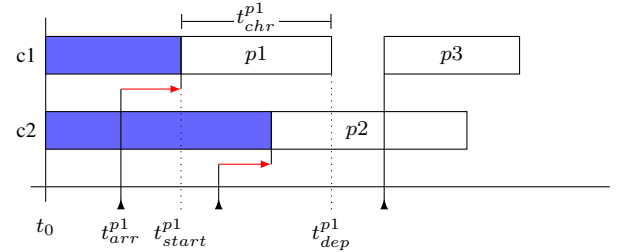
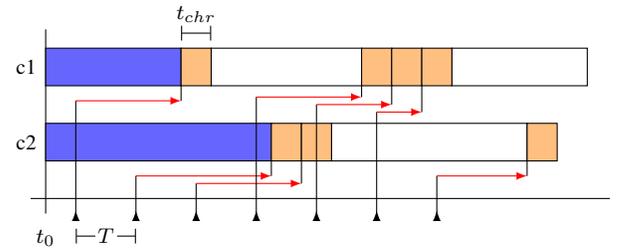
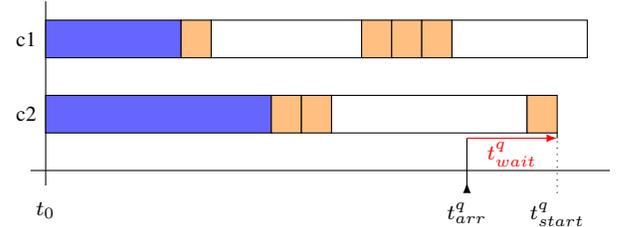
\begin{figure}
	\centering
	\subfloat[Current utilization of charge points (blue) including departure times of the occupying vehicles]{
		\begin{tikzpicture}[every path/.style={>=latex}]
			\draw[-] (0.0, -0.2) -- (0.0, 2.25) {};
			\draw[-] (-0.2,  0.0) -- (7.5, 0.0) {};
			
			\node[anchor=east] at (0, 1.75) {c1}; 
			\node[anchor=east] at (0, 0.75) {c2}; 
			
			\draw[fill=blue!60] (0, 1.5) rectangle +(1.8, 0.5);
			\draw[fill=blue!60] (0, 0.5) rectangle +(3.0, 0.5);

			\node[anchor=north] at (0.0, -0.3) {$t_{0}$};
			\draw[dotted] (1.8, -0.3) -- (1.8, 1.5) {};
			\node[anchor=north] at (1.8, -0.3) {$t_{dep}^{c1}$};
			\draw[dotted] (3.0, -0.3) -- (3.0, 0.5) {};
			\node[anchor=north] at (3.0, -0.3) {$t_{dep}^{c2}$};
			
		\end{tikzpicture}
	}

	\subfloat[Planned charge stops (white) added to the next free charge point after their arrival (triangle). Waiting times (red arrow) may occur if no charge point is free on arrival]{
		\begin{tikzpicture}[every path/.style={>=latex}]
			\draw[-] (0.0, -0.2) -- (0.0, 2.25) {};
			\draw[-] (-0.2,  0.0) -- (7.5, 0.0) {};
			
			\node[anchor=east] at (0, 1.75) {c1}; 
			\node[anchor=east] at (0, 0.75) {c2}; 
			
			\draw[fill=blue!60] (0, 1.5) rectangle +(1.8, 0.5);
			\draw[fill=blue!60] (0, 0.5) rectangle +(3.0, 0.5);
			
			\draw (1.8, 1.5) rectangle node{$p1$} +(2.0, 0.5);
			\draw (3.0, 0.5) rectangle node[xshift=-1ex]{$p2$} +(2.6, 0.5);
			\draw (4.5, 1.5) rectangle node{$p3$} +(1.8, 0.5);
			
			\draw[>-] (1.0, -0.3) -- (1.0, 1.3) {};
			\draw[->, red] (1.0, 1.3) -- (1.8, 1.3) {};
			\draw (1.8, 1.3) -- (1.8, 1.5) {};
			
			\draw[>-] (2.3, -0.3) -- (2.3, 0.3) {};
			\draw[->, red] (2.3, 0.3) -- (3.0, 0.3) {};
			\draw (3.0, 0.3) -- (3.0, 0.5) {};
			
			\draw[>-] (4.5, -0.3) -- (4.5, 1.5) {};
			
			\node[anchor=north] at (0.0, -0.3) {$t_{0}$};
			\node[anchor=north] at (1.0, -0.3) {$t_{arr}^{p1}$};
 			\draw[dotted] (1.8, -0.3) -- (1.8, 1.3) {};
 			\node[anchor=north] at (1.8, -0.3) {$t_{start}^{p1}$};
			\draw[dotted] (3.8, -0.3) -- (3.8, 1.5) {};
			\node[anchor=north] at (3.8, -0.3) {$t_{dep}^{p1}$};
			\draw[|-|] (1.8, 2.2) to node[fill=white,inner sep=1pt,yshift=0.4ex] {$t_{chr}^{p1}$} (3.8, 2.2) {};

		\end{tikzpicture}
	}

	\subfloat[Statistical utilization added as short additional charge stops (orange)]{
		\begin{tikzpicture}[every path/.style={>=latex}]
			\draw[-] (0.0, -0.2) -- (0.0, 2.25) {};
			\draw[-] (-0.2,  0.0) -- (7.5, 0.0) {};
			
			\node[anchor=east] at (0, 1.75) {c1}; 
			\node[anchor=east] at (0, 0.75) {c2}; 
			
			\draw[fill=blue!60] (0, 1.5) rectangle +(1.8, 0.5);
			\draw[fill=blue!60] (0, 0.5) rectangle +(3.0, 0.5);
			
			\draw (2.2, 1.5) rectangle +(2.0, 0.5);
			\draw (3.8, 0.5) rectangle +(2.6, 0.5);
			\draw (5.4, 1.5) rectangle +(1.8, 0.5);
			
			\draw[fill=orange!50] (1.8, 1.5) rectangle +(0.4, 0.5);
			\draw[fill=orange!50] (3.0, 0.5) rectangle +(0.4, 0.5);
			\draw[fill=orange!50] (3.4, 0.5) rectangle +(0.4, 0.5);
			\draw[fill=orange!50] (4.2, 1.5) rectangle +(0.4, 0.5);
			\draw[fill=orange!50] (4.6, 1.5) rectangle +(0.4, 0.5);
			\draw[fill=orange!50] (5.0, 1.5) rectangle +(0.4, 0.5);
			\draw[fill=orange!50] (6.4, 0.5) rectangle +(0.4, 0.5);
			
			\draw[>-] (0.4, -0.3) -- (0.4, 1.3) {};
			\draw[->, red] (0.4, 1.3) -- (1.8, 1.3) {};
			\draw (1.8, 1.3) -- (1.8, 1.5) {};

			\draw[>-] (1.2, -0.3) -- (1.2, 0.3) {};
			\draw[->, red] (1.2, 0.3) -- (3.0, 0.3) {};
			\draw (3.0, 0.3) -- (3.0, 0.5) {};
			
			\draw[>-] (2.0, -0.3) -- (2.0, 0.2) {};
			\draw[->, red] (2.0, 0.2) -- (3.4, 0.2) {};
			\draw (3.4, 0.2) -- (3.4, 0.5) {};

			\draw[>-] (2.8, -0.3) -- (2.8, 1.35) {};
			\draw[->, red] (2.8, 1.35) -- (4.2, 1.35) {};
			\draw (4.2, 1.35) -- (4.2, 1.5) {};
			
			\draw[>-] (3.6, -0.3) -- (3.6, 1.25) {};
			\draw[->, red] (3.6, 1.25) -- (4.6, 1.25) {};
			\draw (4.6, 1.25) -- (4.6, 1.5) {};

			\draw[>-] (4.4, -0.3) -- (4.4, 1.15) {};
			\draw[->, red] (4.4, 1.15) -- (5.0, 1.15) {};
			\draw (5.0, 1.15) -- (5.0, 1.5) {};
			
			\draw[>-] (5.2, -0.3) -- (5.2, 0.3) {};
			\draw[->, red] (5.2, 0.3) -- (6.4, 0.3) {};
			\draw (6.4, 0.3) -- (6.4, 0.5) {};


			\node[anchor=north] at (0.0, -0.3) {$t_{0}$};
			\draw[|-|] (0.4, -0.5) -- node[fill=white,inner sep=1pt] {$T$} (1.2, -0.5) {};
			\draw[|-|] (1.8, 2.2) -- node[yshift=1.6ex] {$t_{chr}$} (2.2, 2.2) {};

		\end{tikzpicture}
	}
	
	\subfloat[Waiting time of the query $t_{wait}^q$ is the difference between the arrival time $t_{arr}^q$ and the charge start time $t_{start}^q$]{
		\begin{tikzpicture}[every path/.style={>=latex}]
			\draw[-] (0.0, -0.2) -- (0.0, 2.25) {};
			\draw[-] (-0.2,  0.0) -- (7.5, 0.0) {};
			
			\node[anchor=east] at (0, 1.75) {c1}; 
			\node[anchor=east] at (0, 0.75) {c2}; 
			
			\draw[fill=blue!60] (0, 1.5) rectangle +(1.8, 0.5);
			\draw[fill=blue!60] (0, 0.5) rectangle +(3.0, 0.5);
			
			\draw (2.2, 1.5) rectangle +(2.0, 0.5);
			\draw (3.8, 0.5) rectangle +(2.6, 0.5);
			\draw (5.4, 1.5) rectangle +(1.8, 0.5);
			
			\draw[fill=orange!50] (1.8, 1.5) rectangle +(0.4, 0.5);
			\draw[fill=orange!50] (3.0, 0.5) rectangle +(0.4, 0.5);
			\draw[fill=orange!50] (3.4, 0.5) rectangle +(0.4, 0.5);
			\draw[fill=orange!50] (4.2, 1.5) rectangle +(0.4, 0.5);
			\draw[fill=orange!50] (4.6, 1.5) rectangle +(0.4, 0.5);
			\draw[fill=orange!50] (5.0, 1.5) rectangle +(0.4, 0.5);
			\draw[fill=orange!50] (6.4, 0.5) rectangle +(0.4, 0.5);

			\draw[>-] (5.6, -0.3) -- (5.6, 0.43) {};
			\draw[->, red] (5.6, 0.43) to node[yshift=-1.34ex] {$t_{wait}^q$} (6.8, 0.43) {};
			\draw[-] (6.8, 0.43) -- (6.8, 0.5) {};
			
			\node[anchor=north] at (0.0, -0.3) {$t_{0}$};
			\node[anchor=north] at (5.6, -0.3) {$t_{arr}^q$};
			\draw[dotted] (6.8, -0.3) -- (6.8, 0.4) {};
			\node[anchor=north] at (6.8, -0.3) {$t_{start}^q$};
		\end{tikzpicture}
	}
	
	\caption{Waiting time estimation example}
	\label{fig:csdbwaitingtimeestimation}
\end{figure}

The database also stores statistical data about the utilization of charging stations in the form of average utilization percentage per hour of a day.
The information can easily be compiled by regularly querying the current utilization of the charging stations.
To account for this statistical utilization in the waiting time estimation, we periodically add short additional charge stops.
The period depends on the charging time $t_{chr}$ of the charge stops, the utilization $u$, and the number of charge points of the charging station $n$:
\begin{equation}
	T = \frac{t_{chr}}{u \cdot n} \text{ .}
\end{equation}
In our experiments, we set $t_{chr}$ to one minute.
This means for a \SI{25}{\percent} utilization at a charging station with two charge points, we would add a charge stop every two minutes.
An example of the waiting time estimation with additional short charge stops can be seen in \cref{fig:csdbwaitingtimeestimation}.

The data in the charging station database can quickly change as additional vehicles announce their planned charge stops.
This means that estimated waiting times, which are queried in the beginning of a vehicles trip, might significantly change by the time the vehicle arrives at the charging stations.
Therefore, to keep the route optimal, we might want to update the route while we are on the trip.
We have define three levels of when this route update could take place:

\emph{Level 1} The route is only calculated once at the beginning and never updated.

\emph{Level 2} The route is updated every time we arrive at a charging station with the option to skip charging at the charging station.

\emph{Level 3} The route may be updated at any point in time on the road, i.e., the charging station database informs the vehicle of any changes in the estimated waiting time.

%

\section{Performance Evaluation}
\label{section:performanceeval}

\subsection{Experimental Setup}

All experiments were run on a 64 core AMD Ryzen Threadripper 3990X CPU at 2.9 GHz and 256 GB of memory.
We implemented our algorithm in C and compiled with GCC 7.5.0 with the highest optimization setting (-O3).

For our experiments, we extracted the road network of Germany from OpenStreetMap excluding very small streets.\footnote{Downloaded from download.geofabrik.de on 2020-02-10. All OSM ways with "highway" tag except for path, steps, elevator, corridor, platform, bridleway, footway, cycleway, pedestrian, proposed, construction, raceway, emergency\_bay, rest\_area, unclassified, residential, living\_street, service, tertiary, tertiary\_link or track.}
It has a total of 4,318,497 nodes, of which many only have decorative purposes to model the shape of the road, only 2,356,510 nodes have more than two edges.
In the preprocessing step, we contracted 4,317,962 (\SI{99.99}{\percent}) of the nodes.

We consider the charging stations from the list provided by German Bundesnetzagentur.\footnote{\url{https://www.bundesnetzagentur.de/DE/Sachgebiete/ElektrizitaetundGas/Unternehmen_Institutionen/HandelundVertrieb/Ladesaeulenkarte/Ladesaeulenkarte_node.html} (visited on 01/04/2020).}
It contains 9,066 charging stations with a total of 21,780 charge points all over Germany with charging powers ranging from \SIrange{2.3}{350}{\kilo\watt}.
Because we are only interested in long distance travel, we used only the fast charging stations of which there are 1,051 with 3,791 charge points.
We precomputed the shortest path trees for these charging stations.

\begin{figure}
	\begin{tikzpicture}
		\begin{axis}[
			xlabel={Time of day (\si{\hour})},
			ylabel={Probability},
			axis x line=bottom, axis y line=left,
			width=8.5cm,height=5cm,
			xtick={0,2,...,24},
			ymin=0,
			xmin=0,
			xmax=24
		]
			\addplot+[mark=none] coordinates {(0,0) (0,0.0267)(9,0.0267) (9,0.05)(12,0.05) (12,0.078)(17,0.078) (17,0.07)(19,0.07) (19,0.016)(24,0.016) (24,0)};
		\end{axis}
	\end{tikzpicture}
	\caption{Departure time distribution (based on distribution of trips on a weekday (Mon--Fri) \cite{pasaoglu2012driving})}
	\label{fig:deptimes}
\end{figure}
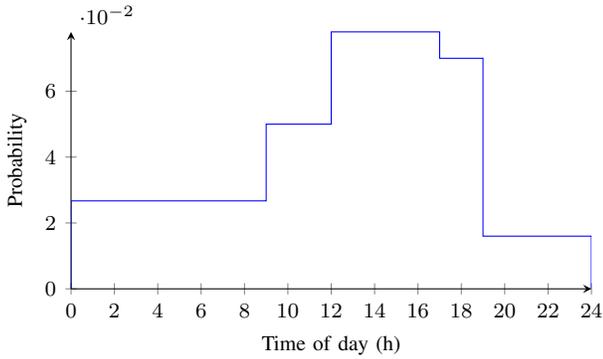

For all tests, we simulated one day with a total of 5,000 electric vehicles doing long distance trips.
Each vehicle was assigned an origin/destination pair with a distance of \SI{500}{\kilo\meter} and a battery capacity ranging from \SIrange{20}{40}{\kilo\watthour}, uniformly distributed, ensuring that recharging on the trip was necessary, often multiple times.
The departure time was selected from a distribution of trips on a weekday (Mon--Fri) in Germany~\cite{pasaoglu2012driving} (see \cref{fig:deptimes}).
We ran each simulation 10 times and averaged the results.

The simulations are computationally expensive, because as part of the charging station routing we have to evaluate all Pareto optimal paths between the origin, destination and charging stations.
By using precomputed shortest path trees for all charging stations, we only have to explore the graph once for the origin and destination nodes to create shortest path trees.
Querying the Pareto optimal paths is then about two orders of magnitude faster than with plain contraction hierarchies, which would explore the graph for each combination again~\cite{schoenberg2019planning}.
However, the query time is still in the order of milliseconds and most queries are between the known locations of the charging stations.
To further improve performance we precomputed all Pareto optimal paths between all combinations of charging stations and stored the costs of the paths in a lookup table.
The lookup table with the cost of all Pareto optimal paths between all 1,051 charging stations has a size of \SI{36.3}{\giga\byte}.

The simulation of 5,000 vehicles without using the \ac{CSDB} took on average \SI{18.6}{\hour} or about \SI{13.4}{\second} per vehicle.
Vehicles that use the \ac{CSDB} levels two or three might have to recompute their route many times, increasing computation times.
The simulation runs mostly in serial on a single thread, apart from exploring the graph to create shortest path trees for origin and destination nodes, which is done in parallel.
For our tests we ran 64 simulations simultaneously, with the precomputed shortest path trees and lookup tables being shared between simulations and held completely in RAM.
The total memory usage of all 64 simulations was between \SIrange{150}{200}{\giga\byte}.

\subsection{Energy Consumption Model}
\label{section:energymodel}

An energy model is required to estimate the energy consumption of an electric vehicle.
The energy consumption is certainly the most important criterion when it comes to optimize the routing of electric vehicles.
For a simplified model, the most important input is the driving speed of the vehicle.
The driving speed impacts the energy consumption due to friction and air drag, which are a function of the speed.
In addition, other energy consuming components of the vehicle need to be considered, e.g., entertainment system, air conditioning, and the head and tail lights.
These components are speed-independent and therefore dominate the energy consumption per \si{\kilo\meter} at lower speeds.

\begin{figure}
	\resizebox {\columnwidth} {!} {
	\begin{tikzpicture}
		\begin{axis}[
			samples=14, domain=20:150, grid,
			xlabel={$v$ (\si{\kilo\meter\per\hour})},
			ylabel={$B$ (\si{\kilo\watthour\per\kilo\meter})},
			axis x line=bottom, axis y line=left,
			ymin=0.1,
			legend pos=north west
		]
			\addplot expression { 0.05 + (x^2 / 90000) + 2/x };
			\addplot table[x=v, y=renault_noneco]{figures/energyconsumption.dat};
			\addplot table[x=v, y=tesla_s85]{figures/energyconsumption.dat};
			
			\legend{Energy model, Renault ZOE, Tesla Model S85}
		\end{axis}
	\end{tikzpicture}
	}
	\caption{Energy model compared to energy consumption of two real electric vehicles}\label{fig:energyconsumptionvsspeed}
\end{figure}
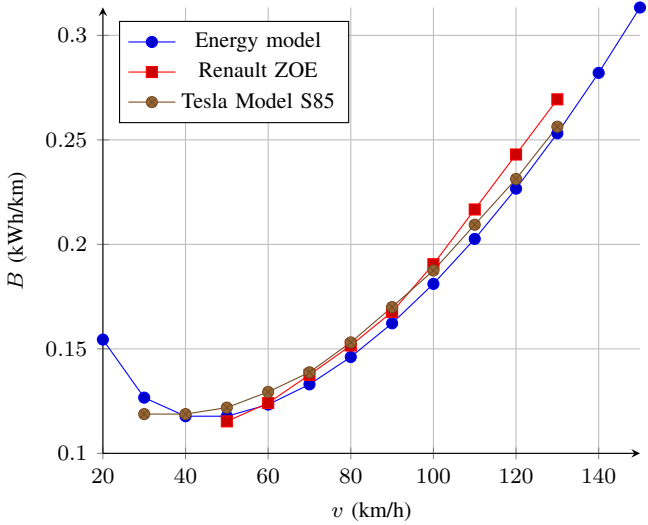

We updated our initial energy model~\cite{schoenberg2019planning} to make it more realistic, by fitting the model to the energy consumption of two real electric vehicles.
We used data available from Renault~\cite{renault2020reichweite}\footnote{Driving range calculator for Renault ZOE with \SI{52}{\kilo\watthour} battery. Eco mode off, temperature \SI{20}{\celsius}, air conditioner and heater off, 15'' wheels} and Tesla~\cite{musk2012model}.
This resulted in the following energy model:
\begin{equation}
B = 0.05 + \frac{v^2}{90000} + \frac{2}{v} \text{ .}
\label{eq:newenergyconsumption}
\end{equation}
The model is plotted in \cref{fig:energyconsumptionvsspeed} together with the energy consumption of the two real electric vehicles.
As can be seen, our energy model very closely matches their energy consumption on a wide range of speeds.
Please note that some factors such as the traffic density or the change in elevation are not incorporated into the model.
However, we believe that the accuracy of the model is sufficient to study and evaluate our proposed approach.

\subsection{Battery Charging Model}
\label{section:chargingmodel}

Traditionally, many authors assumed that the charging speed of a (typical lithium-ion) battery is constant~\cite{alizadeh2014optimized,baum2017consumption}.
However, in reality, this speed is very nonlinear after reaching about \SI{80}{\percent} of the battery's \ac{SOC}.
It actually decreases considerably at that point~\cite{ng2009enhanced}.

Modern lithium-ion batteries are charged with the CC-CV (constant current - constant voltage) charging protocol~\cite{ng2009enhanced,etezadiamoli2010rapidcharge, bashash2011plugin}.
The charging process now follows a two-phase approach.
In the first phase, a constant current approach is used for charging the battery.
During this time, the charge voltages continuously rises.
This process continues until the charge voltage reaches \SI{4.2}{\volt} and the \ac{SOC} is at about \SI{80}{\percent}.
Now the second phase starts using a constant voltage approach to prevent overcharging.
In this phase, the current steadily decreases.
The charging process is assumed to be complete when the current falls below a predefined threshold.
Alternatively, a CP-CV (constant power - constant voltage) protocol can be used.
Here, the charge power is constant in the first phase.
Otherwise, it is very similar to the CC-CV approach.

For our purposes, we use the following battery charging model, which supports both the CC-CV and the CP-CV approach.
We assume that the voltage increase is linear in the first phase and, for simplicity, the current decrease is also linear in the second phase, which is consistent with the literature~\cite{ng2009enhanced}.
For our model, we use the following variables:
The maximum charging power of the charging station is defined as $p_{max}$.
The \ac{SOC} of the battery is defined as $soc$ in the range $0 \leq soc \leq 1$.
In the first phase (constant current/power), the charging voltage increases from $u_{low} = \SI{3.8}{\volt}$ to $u_{high} = \SI{4.2}{\volt}$.
The phase switch happens exactly at $soc = 0.8$.
The maximum current can be calculated as $i_{max} = \dfrac{p_{max}}{u_{high}}$.

Now, the current $i(soc)$ and voltage $u(soc)$ for the CC-CV charging approach can be calculated based on the \ac{SOC} of the battery as
\begin{equation}
i(soc) = \begin{cases}
		i_{max} &\text{for } soc < 0.8\\
		\frac{1 - soc}{0.2} \cdot i_{max} &\text{for } soc \geq 0.8
	\end{cases} \text{ ,}
\label{eq:chargingmodelcurrent}
\end{equation}
\begin{equation}
u(soc) = \begin{cases}
		u_{low} + \frac{soc}{0.8}(u_{high} - u_{low}) &\text{for } soc < 0.8\\
		u_{high} &\text{for } soc \geq 0.8
	\end{cases} \text{ ,}
\label{eq:chargingmodelvoltage}
\end{equation}
\begin{equation}
p_{cc\text{-}cv}(soc) = u(soc) \cdot i(soc) \text{ .}
\label{eq:chargingmodelpowercc}
\end{equation}

Similarly, the power $p_{cp\text{-}cv}(soc)$ can be calculates as
\begin{equation}
p_{cp\text{-}cv}(soc) = \begin{cases}
		p_{max} &\text{for } soc < 0.8\\
		u(soc) \cdot i(soc) &\text{for } soc \geq 0.8
	\end{cases} \text{ .}
\label{eq:chargingmodelpowercp}
\end{equation}
In our algorithm, we estimate the power every second and terminate the charging process when the \ac{SOC} reaches $soc = 0.99$.

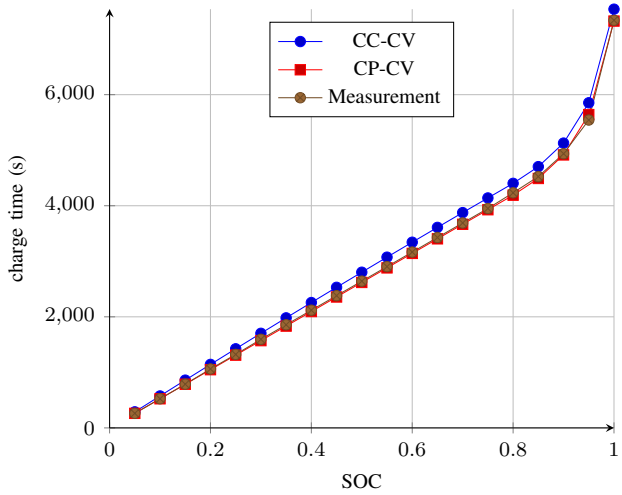
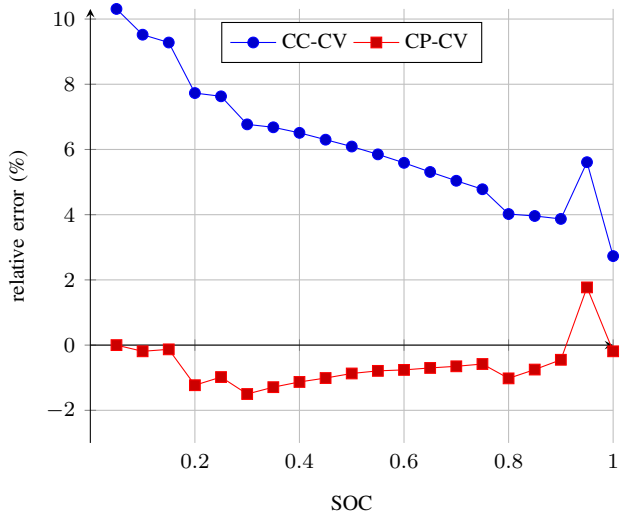
\begin{figure}
	\centering
	\subfloat[Absolute charge time]{
\resizebox {.97\columnwidth} {!} {
	\begin{tikzpicture}
		\begin{axis}[
			grid,
			xlabel={SOC},
			ylabel={charge time (\si{\second})},
			axis x line=bottom, axis y line=left,
			legend style={at={(0.5,0.97)}, anchor=north},
			ymin=0, xmin=0,
		]
			\addplot table[x=SOC,y=CCCV] {figures/cccv_vs_cpcv.dat};
			\addplot table[x=SOC,y=CPCV] {figures/cccv_vs_cpcv.dat};
			\addplot table[x=SOC,y=Measurement] {figures/cccv_vs_cpcv.dat};
			\legend{CC-CV, CP-CV, Measurement}
		\end{axis}
	\end{tikzpicture}
	}
}
	
	\subfloat[Relative error compared to measurement data]{
\resizebox {.97\columnwidth} {!} {
	\begin{tikzpicture}
		\begin{axis}[
			grid,
			xticklabel style={yshift=-3.5em},
			xlabel={SOC},
			ylabel={relative error (\si{\percent})},
			axis x line=center, axis y line=left,
			xlabel style={at={(axis description cs:0.5,-0.1)}, anchor=north},
			legend style={at={(0.5,0.97)}, anchor=north,legend columns=-1},
			ymin=-3, xmin=0	
		]
			\addplot table[x=SOC,y=Error_CCCV] {figures/cccv_vs_cpcv.dat};
			\addplot table[x=SOC,y=Error_CPCV] {figures/cccv_vs_cpcv.dat};
			\legend{CC-CV, CP-CV}
		\end{axis}
	\end{tikzpicture}
	}
}

	\caption{Comparison of CC-CV and CP-CV charging protocols with measurement data.}
	\label{fig:cccvcpccv}
\end{figure}

In a first validation step, we compared our battery charging model with published measurements of an electric vehicle~\cite{zundorf2014electric}.
The results are shown in \cref{fig:cccvcpccv}.
Even though the charging protocol is not mentioned for the measurement data, we can see that the CP-CV approach in our model very closely matches the measurement results.
Actually, the CP-CV approach has a relative error of \SI{\pm2}{\percent}, whereas the CC-CV protocol has a relative error of more than \SI{10}{\percent} at the beginning of the charging process.
We conclude that the vehicle was charged using the CP-CV approach.

\subsection{Experiments}

\begin{figure}
	\begin{tikzpicture}
		\begin{axis}[
			xlabel={CSDB penetration rate (\si{\percent})},
			ylabel={time (\si{\hour})},
			axis x line=bottom, axis y line=left,
			legend style={at={(1.0, 1.0)}, anchor=north east},
			xmin=0, xmax=100,
			ymin=0,
			stack plots=y,
			area style,
		]
			\addplot table[x=PEN_RATE,y=DRIVE_TOT] {figures/csdblevels.dat} \closedcycle;
			\addplot table[x=PEN_RATE,y=CHARGE_TOT] {figures/csdblevels.dat} \closedcycle;
			\addplot table[x=PEN_RATE,y=WAIT_TOT] {figures/csdblevels.dat} \closedcycle;

			\legend{Driving time, Charging time, Waiting time}
		\end{axis}
	\end{tikzpicture}
	\caption{Total travel times of all vehicles for different CSDB penetration rates}
	\label{fig:totaltraveltime_allvehicles}
\end{figure}
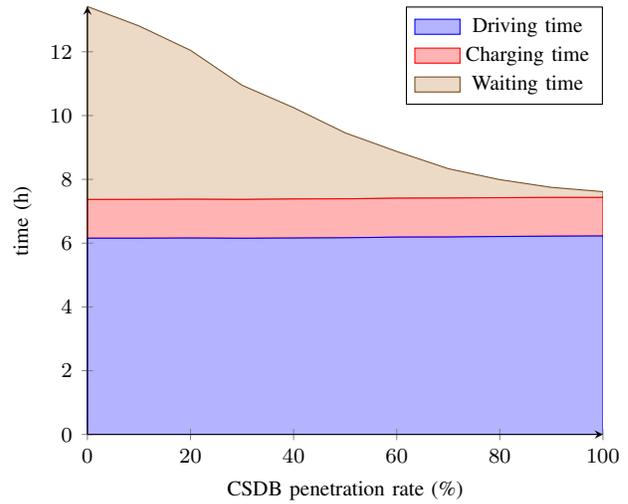
\vspace{0.01em}

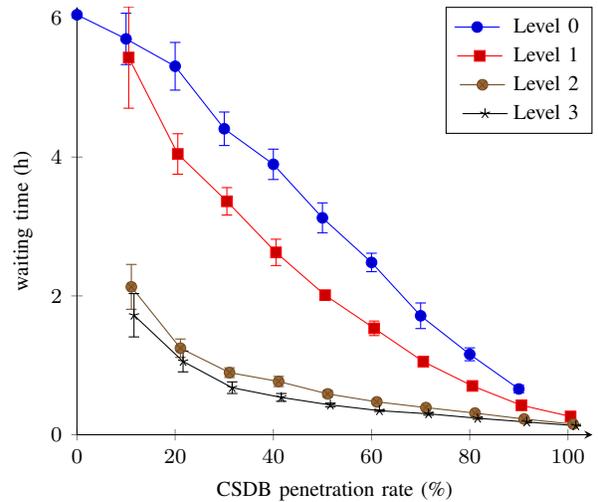
\begin{figure}
	\begin{tikzpicture}
		\begin{axis}[
			xlabel={CSDB penetration rate (\si{\percent})},
			ylabel={waiting time (\si{\hour})},
			axis x line=bottom, axis y line=left,
			legend style={at={(1.0, 1.0)}, anchor=north east},
			ymin=0,
			xmax=105
		]
			\addplot+[error bars/.cd, y dir=both, y explicit] table[x=PEN_RATE,y=WAIT_L0,y error=WAIT_L0_SD] {figures/csdblevels.dat};
			\addplot+[xshift=0.1em, error bars/.cd, y dir=both, y explicit] table[x=PEN_RATE,y=WAIT_L1,y error=WAIT_L1_SD] {figures/csdblevels.dat};
			\addplot+[xshift=0.2em, error bars/.cd, y dir=both, y explicit] table[x=PEN_RATE,y=WAIT_L2,y error=WAIT_L2_SD] {figures/csdblevels.dat};
			\addplot+[xshift=0.3em, error bars/.cd, y dir=both, y explicit] table[x=PEN_RATE,y=WAIT_L3,y error=WAIT_L3_SD] {figures/csdblevels.dat};
			
			\legend{Level 0, Level 1, Level 2, Level 3}
		\end{axis}
	\end{tikzpicture}
	\caption{Waiting times of vehicles with different CSDB levels and CSDB penetration rates}
	\label{fig:waittimes_levels}
\end{figure}

In our first experiment, we examined the influence of the percentage of vehicles that use the \ac{CSDB} (penetration rate) on the total travel time.
We tested penetration rates from \SIrange{0}{100}{\percent} in \SI{10}{\percent} steps.
The vehicles using the \ac{CSDB} were divided equally among the three \ac{CSDB} levels.
As can be seen in \cref{fig:totaltraveltime_allvehicles}, the total travel time of all vehicles is reduced significantly by the use of the \ac{CSDB}.
The reduction is mainly due to decreased waiting times at the charging stations, with only small differences in driving time and charging time.
Without the \ac{CSDB} (\SI{0}{\percent} penetration rate), the average waiting time is 06:03 hours and with all vehicles using it (\SI{100}{\percent} penetration rate), it is reduced to 11 minutes, which is an improvement of about \SI{97}{\percent}.

In \cref{fig:waittimes_levels}, we compare the waiting times for the different \ac{CSDB} levels.
For easier comparison, we refer to not using the \ac{CSDB}, as \ac{CSDB} level 0.
It can be seen, that the \ac{CSDB} level has a big influence on the average waiting times of the vehicles.
At a \SI{10}{\percent} penetration rate, levels 0 and 1 are very close with \SI{5.7}{\hour} and \SI{5.4}{\hour} waiting time, respectively, while levels 2 and 3 are have significantly lower waiting times with \SI{2.1}{\hour} and \SI{1.7}{\hour}, respectively.
This is due to the fact, that the initially planned optimal route for vehicles using \ac{CSDB} level 1, becomes outdated as more and more vehicles announce their planned charge stops.
By updating the route at every charge stop, vehicles using \ac{CSDB} level 2 improved their waiting times substantially.
Vehicles using \ac{CSDB} level 3 improved slightly over level 2, by updating the route while driving.
Obviously, the effect is dependent on the length of the trip and therefore the total travel time.
We tested long distances that likely require multiple charge stops.
For shorter trips with only one charge stop, we would expect the difference to be much smaller.
It can also be observed, that the waiting times of vehicles not using the \ac{CSDB} also improves significantly with higher penetration rates.
The vehicles benefit from the more evenly utilized charging stations with less hot spots causing long waiting times for all vehicles.

\begin{figure}
	\begin{tikzpicture}[spy using outlines={circle, magnification=6, connect spies}]
		\begin{axis}[
			xlabel={Charging station},
			ylabel={Utilization},
			axis x line=bottom, axis y line=left,
			legend style={at={(1.0, 1.0)}, anchor=north east}
		]
			\addplot+[mark=none] table[x expr=\coordindex,y=UTIL_0_PEAK] {figures/csutils.dat};
			\addplot+[mark=none] table[x expr=\coordindex,y=UTIL_100_PEAK] {figures/csutils.dat};

			\coordinate (spypoint) at (axis cs:1,1);
			\coordinate (magnifyglass) at (500, 0.8);
			\legend{\SI{0}{\percent}, \SI{100}{\percent}}
		\end{axis}
		\spy[size=2.5cm] on (spypoint) in node[fill=white] at (magnifyglass);
	\end{tikzpicture}
	\caption{Average utilization of each charging station in peak hours (\SIrange{15}{18}{\hour}) for CSDB penetration rate of \SI{0}{\percent} and \SI{100}{\percent}, sorted by utilization}
	\label{fig:csutils}
\end{figure}
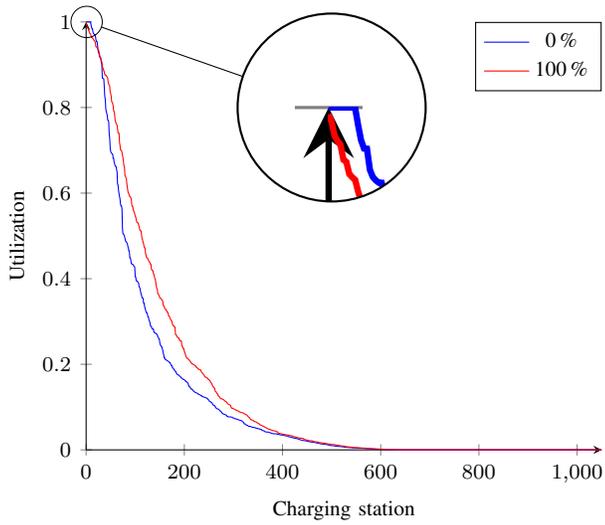

As can be seen in \cref{fig:csutils}, the average utilization during peak hours (\SIrange{15}{18}{\hour}) at a penetration rate of \SI{0}{\percent} is very uneven among the charging stations.
The vast majority of charging stations have a low utilization, only a few hot spots of charging stations are utilized \SI{100}{\percent} during peak hours and contribute to the long waiting times.
At \SI{100}{\percent} penetration rate, the load from the hot spots is shifted to other charging stations, making the utilization of charging stations more even and reducing waiting times significantly.
Even then, the charging stations are very unevenly utilized.
Part of the reason are big differences in charging speed and many charging stations not being close to highways, making them undesirable for long distance travel.
As we focus only on long distance travel in our simulation, the utilization by vehicles doing shorter trips is not considered.

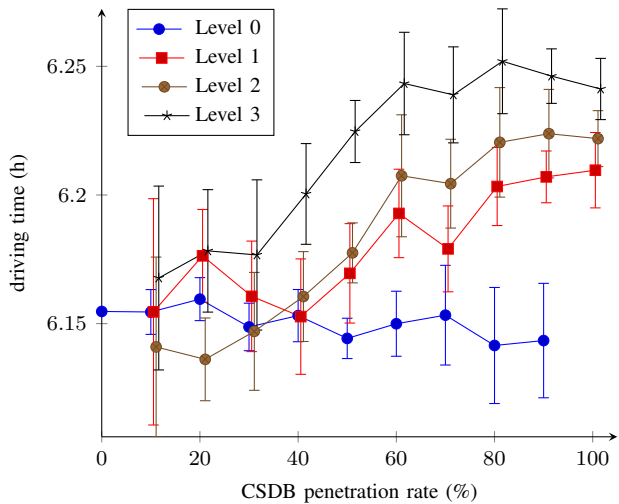
\begin{figure}
	\begin{tikzpicture}
		\begin{axis}[
			xlabel={CSDB penetration rate (\si{\percent})},
			ylabel={driving time (\si{\hour})},
			axis x line=bottom, axis y line=left,
			legend style={at={(0.05, 1.0)}, anchor=north west},
			xmax=105
		]
			\addplot+[error bars/.cd, y dir=both, y explicit] table[x=PEN_RATE,y=DRIVE_L0,y error=DRIVE_L0_SD] {figures/csdblevels.dat};
			\addplot+[xshift=0.1em, error bars/.cd, y dir=both, y explicit] table[x=PEN_RATE,y=DRIVE_L1,y error=DRIVE_L1_SD] {figures/csdblevels.dat};
			\addplot+[xshift=0.2em, error bars/.cd, y dir=both, y explicit] table[x=PEN_RATE,y=DRIVE_L2,y error=DRIVE_L2_SD] {figures/csdblevels.dat};
			\addplot+[xshift=0.3em, error bars/.cd, y dir=both, y explicit] table[x=PEN_RATE,y=DRIVE_L3,y error=DRIVE_L3_SD] {figures/csdblevels.dat};

			\legend{Level 0, Level 1, Level 2, Level 3}
		\end{axis}
	\end{tikzpicture}
	\caption{Driving times of vehicles with different CSDB levels and CSDB penetration rates}
	\label{fig:drivetimes_levels}
\end{figure}

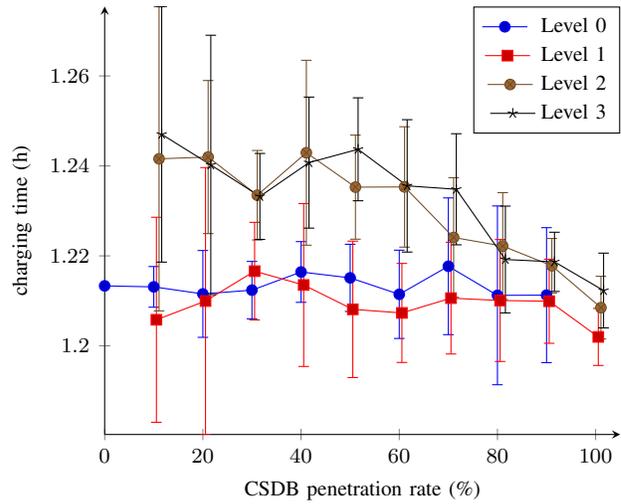
\begin{figure}
	\begin{tikzpicture}
		\begin{axis}[
			xlabel={CSDB penetration rate (\si{\percent})},
			ylabel={charging time (\si{\hour})},
			axis x line=bottom, axis y line=left,
			legend style={at={(1.0, 1.0)}, anchor=north east},
			xmax=105
		]
			\addplot+[error bars/.cd, y dir=both, y explicit] table[x=PEN_RATE,y=CHARGE_L0,y error=CHARGE_L0_SD] {figures/csdblevels.dat};
			\addplot+[xshift=0.1em, error bars/.cd, y dir=both, y explicit] table[x=PEN_RATE,y=CHARGE_L1,y error=CHARGE_L1_SD] {figures/csdblevels.dat};
			\addplot+[xshift=0.2em, error bars/.cd, y dir=both, y explicit] table[x=PEN_RATE,y=CHARGE_L2,y error=CHARGE_L2_SD] {figures/csdblevels.dat};
			\addplot+[xshift=0.3em, error bars/.cd, y dir=both, y explicit] table[x=PEN_RATE,y=CHARGE_L3,y error=CHARGE_L3_SD] {figures/csdblevels.dat};
			
			\legend{Level 0, Level 1, Level 2, Level 3}
		\end{axis}
	\end{tikzpicture}
	\caption{Charging times of vehicles with different CSDB levels and CSDB penetration rates}
	\label{fig:chargetimes_levels}
\end{figure}

The driving times and charging times are also slightly influenced by using the \ac{CSDB}.
As can be seen in \cref{fig:drivetimes_levels}, the average driving time correlates with a higher penetration rate and a higher \ac{CSDB} level.
As more vehicles use the \ac{CSDB}, the vehicles have more information about planned charge stops of other vehicles in the future and are therefore more likely to drive detours to alternative charging stations with lower waiting times.
This is especially true for vehicles using higher \ac{CSDB} levels, as they can update their route, which may lead to additional detours based on planned charge stops that have been announced in the meantime.
This is not a problem, as the additional driving time is only small and is more than compensated by the saved waiting time.
The average charging times can be seen in \cref{fig:chargetimes_levels}.
Vehicles using \ac{CSDB} levels 2 and 3 have a slightly higher charging time at low penetration rates.
This is caused by the vehicles changing their selected charging stations to less optimal ones, in case there are other unexpected vehicles at the selected charging station.
As more vehicles use the \ac{CSDB}, encountering unexpected vehicles becomes less likely.

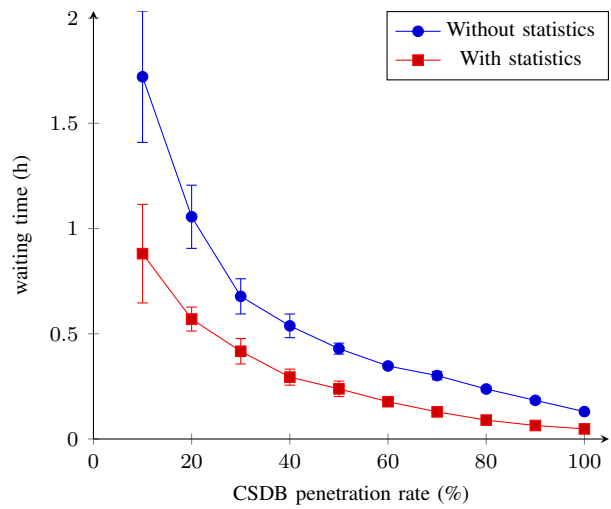
\begin{figure}
	\begin{tikzpicture}
		\begin{axis}[
			xlabel={CSDB penetration rate (\si{\percent})},
			ylabel={waiting time (\si{\hour})},
			axis x line=bottom, axis y line=left,
			legend style={at={(1.0, 1.0)}, anchor=north east},
			xmin=0, xmax=105,
			ymin=0,
		]
			\addplot+[xshift=0.0em, error bars/.cd, y dir=both, y explicit] table[x=PEN_RATE,y=WAIT_L3,y error=WAIT_L3_SD] {figures/csdblevels.dat};
			\addplot+[xshift=0.0em, error bars/.cd, y dir=both, y explicit] table[x=PEN_RATE,y=WAIT_L3,y error=WAIT_L3_SD] {figures/csdblevelsstats.dat};

			\legend{Without statistics, With statistics}
		\end{axis}
	\end{tikzpicture}
	\caption{Waiting times compared with and without using statistics on CSDB Level 3}
	\label{fig:waittimes_statistics}
\end{figure}

In our second experiment, we compared using statistical data about the charging station utilization with only using the current utilization and planned charge stops.
We generated the statistical data from the charging station utilization of simulation runs of the first experiment.
In \cref{fig:waittimes_statistics}, we compare the average waiting times with and without using statistics for vehicles using \ac{CSDB} Level 3.
It can be seen that using statistical data approximately halves the waiting times.

Interestingly, it not only improves the waiting times for low penetration rates, where many vehicles are not accounted for by the \ac{CSDB}, but also for high penetration rates including \SI{100}{\percent}, where all vehicles announce their planned charge stops.
When a vehicle initially plans its trip at departure time, the \ac{CSDB} will only know about planned charge stops of vehicles that have already departed.
Even though the vehicle can update its route while driving, by the time it becomes known that it is on a suboptimal path, it might already be too late to change it.
It is therefore beneficial to account for vehicles departing in the future by using statistical data.

\begin{figure}
	\centering
	
	\resizebox {0.95\columnwidth} {!} {
	\subfloat[Total travel time for different CSDB penetration rates]{
	\begin{tikzpicture}
		\begin{axis}[
			xlabel={CSDB penetration rate (\si{\percent})},
			ylabel={total travel time (\si{\hour})},
			axis x line=bottom, axis y line=left,
			legend style={at={(1.0, 0.57)}, anchor=east},
			xmin=0, xmax=105
		]
 			\addplot+[xshift=0.0em, error bars/.cd, y dir=both, y explicit] table[x=PEN_RATE,y=AA,y error=AA_SD] {figures/strategies.dat};
 			\addplot+[xshift=0.1em, error bars/.cd, y dir=both, y explicit] table[x=PEN_RATE,y=MA,y error=MA_SD] {figures/strategies.dat};
 			\addplot+[xshift=0.2em, error bars/.cd, y dir=both, y explicit] table[x=PEN_RATE,y=8A,y error=8A_SD] {figures/strategies.dat};
			\addplot+[xshift=0.3em, error bars/.cd, y dir=both, y explicit] table[x=PEN_RATE,y=FA,y error=FA_SD] {figures/strategies.dat};
 			\addplot+[xshift=0.4em, error bars/.cd, y dir=both, y explicit] table[x=PEN_RATE,y=AF,y error=AF_SD] {figures/strategies.dat};
 			\addplot+[xshift=0.5em, error bars/.cd, y dir=both, y explicit] table[x=PEN_RATE,y=AE,y error=AE_SD] {figures/strategies.dat};
			
			\legend{Adaptive, Min charge, \SI{80}{\percent} charge, Full charge, Fastest route, Economic route}
		\end{axis}
	\end{tikzpicture}
	}
	}
\vspace{0.01em}

	\resizebox {0.95\columnwidth} {!} {
	\subfloat[Total travel time composition at \SI{30}{\percent} and \SI{70}{\percent} penetration rate]{
	\begin{tikzpicture}[
		every axis/.style={
			ybar stacked,
			symbolic x coords={AA, MA, 8A, FA, AF, AE},
			xtick=data,
			ymin=0,ymax=10.5
		}
	]
		\begin{axis}[bar shift=-6pt,
			axis line style={draw=none},
			tick style={draw=none},
			yticklabels={,,},
			x tick label style={xshift=-6pt,rotate=90,anchor=east},
			xticklabels={\SI{30}{\percent}, \SI{30}{\percent}, \SI{30}{\percent}, 
						\SI{30}{\percent}, \SI{30}{\percent}, \SI{30}{\percent}}
		]
 			\addplot table[x=STRATEGY,y=DRIVE_30]{figures/strategies_30_70.dat};
 			\addplot table[x=STRATEGY,y=CHARGE_30]{figures/strategies_30_70.dat};
 			\addplot table[x=STRATEGY,y=WAIT_30]{figures/strategies_30_70.dat};

		\end{axis}
		\begin{axis}[bar shift=+6pt,
			legend style={at={(0.5,1.15)}, anchor=north,legend columns=-1},
			ylabel={total travel time (\si{\hour})},
			x tick label style={xshift=+6pt,rotate=90,anchor=east},
			xticklabels={\SI{70}{\percent}, \SI{70}{\percent}, \SI{70}{\percent}, 
						\SI{70}{\percent}, \SI{70}{\percent}, \SI{70}{\percent}}
		]
			\addplot table[x=STRATEGY,y=DRIVE_70]{figures/strategies_30_70.dat};
			\addplot table[x=STRATEGY,y=CHARGE_70]{figures/strategies_30_70.dat};
			\addplot table[x=STRATEGY,y=WAIT_70]{figures/strategies_30_70.dat};
			\legend{Driving time, Charging time, Waiting time}
		\end{axis}
		
		\begin{axis}[bar shift=+6pt,
			axis line style={draw=none},
			tick style={draw=none},
			yticklabels={,,},
			x tick label style={yshift=-30pt,rotate=45,anchor=east},
			xticklabels={Adaptive, Min charge, \SI{80}{\percent} charge, Full charge, Fastest route, Economic route},
		]
			\addplot coordinates {(AA,0) (MA,0) (8A,0) (FA,0) (AF,0) (AE,0)};
		\end{axis}
	\end{tikzpicture}
	}
	}
	\caption{Comparison of adaptive charging and routing strategy to other strategies (CSDB level 3 with statistics)}
	\label{fig:totaltraveltimes_strategies}
\end{figure}
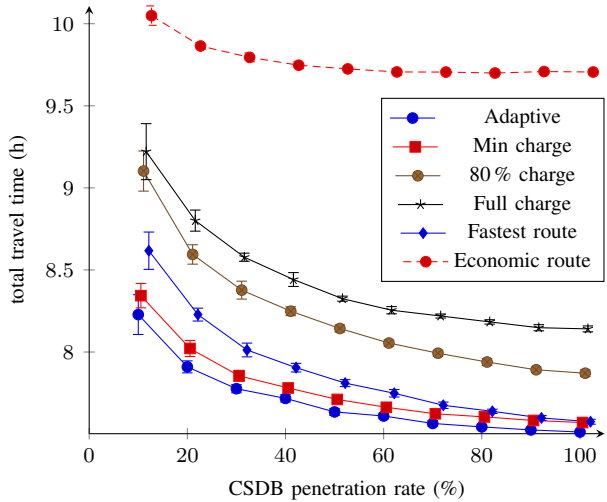
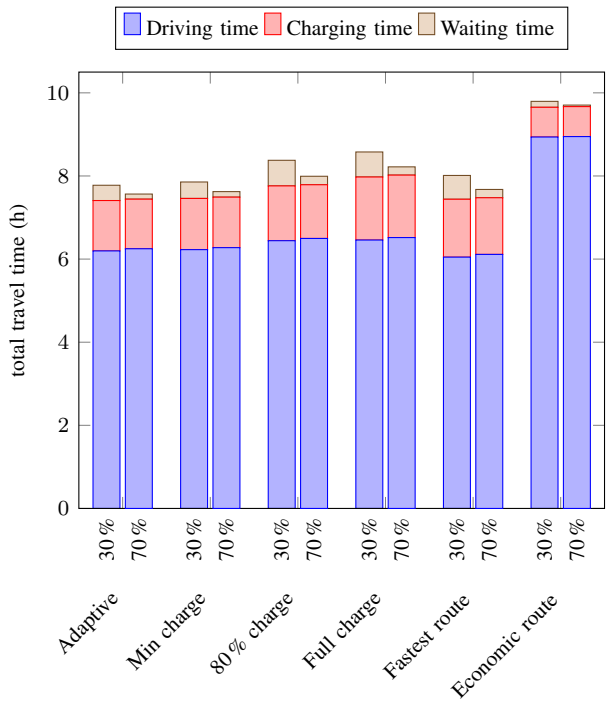

In our third experiment, we compared our adaptive charging and routing strategy with alternative strategies often found in literature.
For each strategy, we ran a set of simulations with vehicles using \ac{CSDB} level 3 and statistics at penetration rates \SIrange{10}{100}{\percent} in \SI{10}{\percent} steps.
In \cref{fig:totaltraveltimes_strategies}, the total travel time of the strategies over all penetration rates and the composition of the total travel time for the penetration rates \SI{30}{\percent} and \SI{70}{\percent} are shown.

We can see that using the adaptive strategy results in the least total travel time in all cases.
It is very closely followed by always doing a minimum charge, just to get to the next charging station (min charge).
The adaptive charging strategy only has an advantage over min charge in cases where the first charge is at a fast charging station and then a second charge is at a slower charging station.
As there are plenty fast charging stations available, this is a rare case.
Always doing an \SI{80}{\percent} charge or always doing a full charge, causes significantly more charging time, because more energy then necessary is being charged.
This makes choosing a fast charging station more important to the algorithm, which results in additional driving time due to detours and additional waiting time due to higher utilizations at fast charging stations.

The adaptive routing strategy selects the route out of the set of Pareto optimal routes that minimizes driving, charging and waiting time.
We compared this with always choosing the fastest route and always choosing the most economic route.
Always choosing the fastest route reduces the driving time, but leads to more charging time due to more energy consumption, which in turn leads to more waiting time due to higher charging station utilization.
The opposite can be observed when always choosing the most economic route.
The charging time is reduced significantly due to less energy consumption, which also leads to a very short waiting time due to lower charging station utilization, but the driving time is significantly longer, making it the strategy with the highest total travel time.

In all cases, we can see that a higher penetration rate of the \ac{CSDB} reduces the waiting time and the total travel time.

%

\acresetall
\section{Conclusion}
We presented an approach to minimize waiting times at charging stations for long distance trips with electric vehicles by announcing planned charge stops to a central \ac{CSDB}.
We integrated the waiting time estimates of the \ac{CSDB} into our adaptive charging station routing approach to minimize the total travel time of electric vehicles.
In our evaluation, we considered the map of Germany and simulated one day with a large number of vehicles doing long distance trips.
The existing heterogeneous charging infrastructure with its differences in charge power and number of charge points was used together with a realistic non-linear charge model.
We showed that the utilization of charging stations is very uneven and can cause long waiting times, but that by using the \ac{CSDB}, average waiting times can be reduced by up to \SI{97}{\percent}.
As the waiting time estimates can become outdated after a while, the route of long distance trips should be updated at least at every charge stop.
Updating the route while driving further improves the waiting time.
By using statistical data about the utilization of the charging stations derived from historical data, we could additionally reduce the waiting time by about half.
Furthermore, we compared using the \ac{CSDB} with our adaptive charging and routing strategies to other strategies often found in literature.
We could clearly show, that while using the \ac{CSDB} is beneficial to reducing waiting times in all cases, the combination with our adaptive strategies provides the best total travel times.

In future work, we want to better take human behavior and individual preferences into account.
Human drivers might be frustrated by frequent route changes or need to rest some time after a long drive.
We also want to look into using the \ac{CSDB} and adaptive strategies for short distance trips where recharging is not necessary to reach the destination, but the driver simply wants to recharge the vehicle and could either stop at a charging station on the way or charge near his destination, which might entail additional walking time.
In this context we want to compare having many slow charging stations to having few fast charging stations.

%

\printbibliography

\end{document}